\documentclass{aa}
\usepackage{graphicx}
\usepackage{natbib}
\usepackage{color}
\usepackage{rotating,color}
\usepackage{supertabular}
\usepackage{longtable}

\begin{document}

\def\kms{~km~s$^{-1}${}}
\def\gax    {{_>\atop^{\sim}}}
\def\deg{$^{\circ}$}
\def\msyr{~M$_{\odot}$~yr$^{-1}$}
\def\mspc{~M$_{\odot}$~pc$^{-2}$}
\def\dexkpc{~dex~kpc$^{-1}$}

\def\ms{M$_{\odot}$}
\def\zs{Z$_{\odot}$}
\newcommand{\dkp}{dex kpc$^{-1}$}

\title{Milky Way vs Andromeda: a tale of two disks}

\author{ J. Yin\inst{1,5}, J.L. Hou\inst{1}, N. Prantzos\inst{2}, S. Boissier\inst{3},
R.X. Chang\inst{1}, S.Y. Shen\inst{1}, B. Zhang\inst{4}
        }

\authorrunning{Yin et al. }

\titlerunning{Evolution of Milky Way and Andromeda}

\institute{ Key Laboratory for Research in Galaxies and Cosmology,
Shanghai Astronomical Observatory, CAS, 80 Nandan Road, Shanghai,
200030, China  \and CNRS, UMR7095, Institut d'Astrophysique de
Paris, F-75014, Paris, France \and Laboratoire d'Astrophysique de
Marseille, BP8, Traverse du Siphon, 13376 Marseille, Cedex 12,
France \and Department of Physics, Hebei Normal University, 113
Yuhua Dong Road, Shijiazhuang, 050016, China \and
 Graduate School, the Chinese Academy of Sciences, Beijing,
100039, China }
\date{Submitted 2009}

\abstract{} {We study the chemical evolution of the disks of the
Milky Way (MW) and of Andromeda (M31), in order to reveal common
points and differences between the two major galaxies of the Local
group.} {We use a large set of observational data for M31, including
recent observations of the Star Formation Rate (SFR) and gas
profiles, as well as stellar metallicity distributions along its
disk. We show that, when expressed in terms of the corresponding
disk scale lengths, the observed radial profiles of MW and M31
exhibit interesting similarities, suggesting the possibility of a
description within a common framework.} {We find that the profiles
of stars, gas fraction and metallicity of the two galaxies, as well
as most of their global properties, are well described by our model,
provided the star formation efficiency in M31 disk is twice as large
as in the MW. We show that the star formation rate profile of M31
cannot be fitted with any form of the Kennicutt-Schmidt law (KS Law)
for star formation. We attribute those discrepancies to the fact
that M31 has undergone a more active star formation history, even in
the recent past, as suggested by observations of a ``head-on''
collision with the neighboring M32 galaxy about 200 Myr ago.} {The
MW has most probably undergone a quiescent secular evolution, making
possible a fairly successful description with a simple model. If M31
is more typical of spiral galaxies, as recently suggested by Hammer
et al. (2007), more complex models, involving galaxy interactions,
will be required for the description of spirals.}


\maketitle

\section{Introduction}

Substantial progress has been made in the last decade or so in our
understanding of the evolution of disk galaxies in a cosmological
context, both with analytical models (Mo, Mao \& White 1998;
Efstathiou 2000) and with numerical simulations, getting to higher
resolution and including more and more physical ingredients: dark
matter profiles and assembly, baryon accretion histories, gas
cooling and supernova feedback, multi-phase interstellar medium
(e.g. Cole et al. 2000; Samland \& Gerhard 2003; Monaco 2004; Kang
et al. 2005; Heller et al. 2007; Guo \& White 2008; Ro\v{s}kar et
al. 2008). However, the complexity of the relevant baryonic physics
is not yet fully understood and no satisfactory disk model exists at
present, although the most recent simulations with high resolution
and improved stellar feedback start producing  disks resembling the
observed ones (Governato et al. 2007; Mayer et al. 2008).

Because of those difficulties, the simple phenomenological models
developed in the 80ies and 90ies may still be of considerable help.
In those models, the complex processes related to disk formation
through gas accretion (i.e. merging histories, gas cooling, stellar
feedback, etc.) are simply described by an analytical infall law.
Such models do not construct the galaxy ``ab initio'' but rely on
the observed present-day features of a galaxy in order to infer its
past history, thus they have been characterized as ``backwards''
models (e.g. Ferreras \& Silk 2001). They have been widely used in
studies of the chemical evolution of the Milky Way (e.g. Matteucci
\& Francois 1989; Prantzos \& Aubert 1995; Chiappini et al. 1997;
Prantzos \& Silk 1998; Boissier \& Prantzos 1999; Chang et al. 1999,
2002; Cescutti et al. 2007 etc.) allowing important progress towards
our understanding of our Galaxy's evolution. Indeed, some
convergence has been reached among the various groups concerning
e.g. i) the necessity of substantial infall on long time scales (to
explain the local G-dwarf metallicity distribution and the small
degree of astration of deuterium); ii) the necessity of  radially
varying time scales for the infall and the star formation rate (to
obtain the observed profiles of metallicity, gas fraction and
colors), and iii) no need for varying IMF or strong galactic winds.

As one of the three disk galaxies in the Local Group, Andromeda
(M31, or NGC224) provides a unique opportunity for testing theories
of galaxy formation and evolution (Klypin et al. 2002; Widrow et al.
2003; Widrow \& Dubinski 2005; Geehan et al. 2006; Tamm et al. 2007;
Tempel et al. 2007). The wealth of available data can be used to
constrain models of the evolution of the disk, bulge and halo of
M31. However, due to its size, proximity and big bulge, most of the
work has been done on the stellar and kinematic properties of M31
halo and globular clusters (Beasley et al. 2004; Burstein et al.
2004; Durrell et al. 2004; Chapman et al. 2006; Font et al. 2008;
Koch et al. 2008; Lee et al. 2008), outer disk (Ibata et al. 2005;
Irwin et al. 2005; Worthey et al 2005; Brown et al. 2006,2007,2008;
Richardson et al. 2008) and the central bulge (Jacoby \& Ciardullo
1999; Salow \& Statler 2004; Sarajedini \& Jablonka 2005; Davidge et
al. 2006; Olsen et al. 2006).

Star formation (SF) histories in various regions of the M31 disk and
halo have also been measured with the Hubble Space Telescope (HST)
and ground based large telescopes. For example, Williams (2003a,
2003b) has measured the star formation history in several regions of
the M31 disk from the KPNO/CTIO Local Group Survey and  found that
the total mean star formation rate for the disk is about 1 \ms
yr$^{-1}$. With deep HST photometry, Bellazzini et al. (2003) and
Worthey et al. (2005) have studied the stellar abundance
distributions and star formation history in many locations of the
disk (see also Ferguson \& Johnson 2001; Sarajedini \& van Duyne
2001; Williams 2002; Brown et al. 2006; Olsen et al. 2006). Those
observations have revealed that the M31 disk has a mean disk age
around 6-8 Gyr and mean metallicity of [Fe/H]$\sim-0.2$, albeit with
substantial spread in both cases.

Compared with the Milky Way, M31 appears to have been more active in
the past, although its current star formation rate is smaller than
that of our Galaxy. Based on a survey of spiral properties, Hammer
et al. (2007) suggested that the Milky Way is a rather quiescent
galaxy, untypical of its class, while the M31 may be closer to a
typical spiral. Using detailed two components (disk+halo) chemical
evolution models, Renda et al. (2005) have compared some chemical
properties between M31 and MW disk/halo, and conclude that  M31 must
have a higher star formation efficiency and/or shorter infall time
scale. Deep photometry of the M31 halo shows that it hosts
populations of old and metal-poor stars, along with younger and of
higher metallicity ones, pointing to a prolonged period of active
merging. (Brown 2009 and references therein). The two ring-like
structures observed in M31 (Block et al. 2006) are interpreted as
the result of a recent ($<$200 Myr ago) collision with a companion
galaxy (Block et al. 2006) and give support to the idea of recent
merging activity of M31.

In this paper, we attempt a comparative study of the chemical
evolution of MW and M31, by constraining our model with a more
extended data set than in any previous work. Our data include global
properties and radial profiles of gas, stars, gas fraction, star
formation rate, and oxygen abundances, as well as stellar
metallicity distributions along the disk of M31 (Sec. 2). We find
that, when the radial profiles are expressed in terms of the
corresponding scale lengths of the stellar disks, the MW and M31
present some interesting similarities (Sec. 2.4) encouraging us to
adopt a single phenomenological model for the description of both
galaxies (Sec. 3). The model describes fairly well all of the key
properties of MW and most of M31 (Sec. 4), provided the star
formation efficiency is twice as large in the latter case. We
discuss the successes and failures of the model, and we compare to
previous work in Sec. 5. Sec. 6 summarizes our results.

\section{Observational Properties }

\subsection{Stellar disks: Scale lengths and masses }

The stellar disks of Milky Way and M31 are well described by
exponential surface density profiles, given by:
\begin{equation}\label{eq:sigma}
    \Sigma_{*}(r,t_g)=\Sigma(r_0,t_g)e^{-(r-r_0)/r_d}
\end{equation}
where $r_d$ is the disk scale length and $\Sigma(r_0,t_g)$ is the
local surface density at some distance $r_0$ from the galactic
center at the present time $t_g$=13.5 Gyr. In the case of the Milky
Way, the reference distance is obviously Sun's distance of
$r_0=R_{\odot MW}$=8 kpc, where the local stellar surface density is
evaluated to $\Sigma(r_0,t_g)$=37 \mspc \ (Flynn et al. 2006). The
total stellar mass of the disk is then given by
\begin{equation}\label{eq:diskmass}
M_d \ = \ \int_{r_b}^{r_2}   2 \pi \ r \ \Sigma_{*}(r,t_g) dr
\end{equation}
where $r_b$=2.5 kpc  is the bulge radius and $r_2$ the outer disk radius.

Observed disk scale lengths  are obtained from measurements  of
surface brightness profiles in various wavelength bands and they are
wavelength dependent. $B$ band scale length reflects mostly the SFR
profile in the past $\sim$ 1 Gyr, while  $K$ or $R$ scale lengths
reflect the total stellar population, cumulated over the age of the
disk.

For the Milky Way disk, we adopt the mean value of $r_d = 2.3\pm0.6$
kpc (from measurements in the $R$ or $I$ bands), derived from the
compilation of Hammer et al. (2007). The total  mass up to 15 kpc is
then $\sim$~3$\times$10$^{10}$ \ms.

By adding $\sim$0.7$\times$10$^{10}$ \ms~for the gaseous disk mass
(as estimated in the next section) one gets a total baryonic disk
mass of $\sim$3.7$\times$10$^{10}$ \ms, in good agreement with mass
models of the Milky Way (e.g. Dehnen \& Binney 1998, Naab and
Ostriker 2006).

Based on the observed disk surface brightness of M31, Walterbos \&
Kennicutt (1987, 1988) have measured the disk scale length in
different wavelengths. They obtained $r_d$ = 6.8, 5.8, 5.3, and 5.2
kpc in the $U,B,V,$ and $R$ bands, respectively. Recently, Worthey
et al.(2005) obtained 5.6 kpc in the $I$ band, while for $K$ band
Hiromoto et al. (1983) find  $r_d$ = 4.2 kpc. With the IRAC on board
the SPITZER space telescope, Barmby et al. (2006) measured, for the
first time, the mid infrared surface brightness profile of M31 and
found a scale length of 6.08 kpc in the $L$ band. Note that
different authors adopt different distance scale of M31. In Table 1,
we list all the available observed disk scale lengths and scale them
to the same distance of 785 kpc (McConnachie et al. 2005). Overall,
the values are consistent for different bands, except for the
shorter wavelengths which are likely to be affected by dust
extinction. In this paper we adopt an averaged value from four
observed values from three bands ($R, I, K$), which is $r_d$=5.5
kpc. This value is within the range of $r_d$=5.8$\pm$0.4 kpc found
in Hammer et al. (2007).

The total mass of M31 disk is obtained through observational data
and  mass models. In their disk-bulge-halo model, Widrow et al.
(2003) find  that their best model requires the M31 disk mass (stars
+ gas) to be about 7 $\times$ 10$^{10}$ \ms. Recent mass model of
Geehan et al. (2006) also gives a similar disk mass value  $\sim$7.2
$\times$ 10$^{10}$ \ms, by adopting a disk mass-to-light ratio of
3.3. In this paper, we adopt then the M31 total disk mass to be
M$_{tot}$ = 7$\times$10$^{10}$\ms. By subtracting $\sim$6 $\times$
10$^9$ \ms \ for the gas (see next section) we obtain a total disk
stellar mass of 5.9 $\times$ 10$^{10}$ \ms \ for Andromeda.

In  summary, the M31 disk is about 2 times as massive and 2.4 times
as large as the Milky Way disk.

\begin{table*} [!t]

\begin{center}

{\bf Table 1} Observation of Milky Way and M31 (re-scaled to 785
kpc)
\begin{tabular}{lllll}
     \hline  \hline
     Observable                  & Milky Way     & reference  & M31       & reference \\
    \hline \hline
    \textbf{Global properties} \\
    \hline
     Type                         & SbcI-II      & 1   & SbI-II           & 1\\
     $K$-band                       & $M_K=-24.02$ & 2   & $M_K=-24.70$     & 3 \\
     Total luminosity             &              &     &                  &   \\
      \qquad $L_B$ ($10^{10}L_{B\odot}$) &  1.8  & 4   & 3.3              & 5 \\
      \qquad $L_V$ ($10^{10}L_{V\odot}$) &  2.1  & 6   & 2.6$\sim$2.7     & 1 \\
      \qquad $L_K$ ($10^{10}L_{K\odot}$) &  5.5  & 7   & 6 $\sim$ 12$^a$  & 8 \\
     Total color ($B-V$)                 &  0.84 & 4   & 0.81             & 9\\
     Mass                                &       &     &                  &  \\
      \qquad Total ($10^{10}$\ms)        & 40-55 & 10, 11 & $107-140$     & 12\\
      \qquad Visible ($10^{10}$\ms)      & 5.0   & 3      & $5.9-8.7$     & 5\\
     Rotational Curve (\kms)             &       &        &                &  \\
      \qquad Flat velocity               & 220   & 1      & 226            & 13 \\
    \hline \hline
     \textbf{Bulge}                      &       &        &                &   \\
    \hline
      \qquad Stellar mass ($10^{10}$\ms) & 1-2   & 14     & 3.2            & 15  \\
      \qquad Effective radius (kpc)      & 2.5   & 1      & 2.6            & 1   \\
    \hline \hline
     \textbf{Disk}          &                    &        &                & \\
    \hline
     Scale length (kpc)       &                    &        &                &  \\
      \qquad $U$              &                    &        &  7.7           & 9 \\
      \qquad $B$              & $4.0\sim5.0$       & 7      &  6.6           & 9 \\
      \qquad $V$              & $2.5\sim3.5$       & 6, 16  &  6.0           & 9 \\
      \qquad $R$              &  2.3               & 3      &  5.9           & 9 \\
      \qquad $I$              &                    &        & 5.7            & 17 \\
      \qquad $K$              & $2.3\sim2.8$       & 18     & 4.8           & 19 \\
      \qquad $L$              &                    &        & 6.08          & 8 \\
     Total SFR (\ms $yr^{-1}$)  & $\sim$1-5        & 20,21   & 0.35 - 1.0  & 1, 8, 21, 22\\
     Infall rate (\ms $yr^{-1}$)& $0.5\sim5$     & 21, 23 &               & \\
     Total mass of          &                    &        &                  & \\
      \qquad disk($10^{10}$\ms) & 3.5 $^b$       & 24     & $\sim$7       & 15, 25, 26\\
      \qquad star($10^{10}$\ms) & 3.0 $^c$       & 16     & $\sim$6   & 5\\
      \qquad gas($10^{10}$\ms)  & $\sim$0.7  & 27, 28 & $\sim$0.6             & 29\\
      \qquad HI($10^{10}$\ms)   & 0.4            & 1      & $\sim$0.5   & 1, 29, 30\\
      \qquad H$_2$($10^{10}$\ms)& 0.11           & 31     & $\sim$0.02-0.04 & 29, 30, 31\\
     Gas fraction               & $\sim$0.15-0.2 & 20, This paper & $\sim$ 0.09  & This paper \\
     Abundance gradient         &                  &            &                    & \\
      \qquad [O/H] (\dexkpc)  & $-0.04 \sim -0.07$ & 32, 33, 34 & $-0.018 \sim -0.027$ & 35, 36\\
    Color gradient                &                &            &                    & \\
      \qquad $B-V$ (mag kpc$^{-1}$) &              &            & 0.016              & 9\\
    \hline
    \hline
\end{tabular} \\
\end{center}
Note:\\
a: assuming M/$L_K$ = 1.15 (M/$L_K)_{\odot}$ and  M31 mass is taken
to be (7-14)$\times$ $10^{10}$ $M_\odot$ \\
b: derived based on the disk scale length $r_d$=2.3 kpc and total
disk surface density at the solar neighborhood $\Sigma_{tot}$ =
50 \ms $pc^{-2}$ \\
c: derived based on the stellar disk scale length $r_d$=2.3 kpc and
stellar surface density at the solar neighborhood $\Sigma_{*}$ = 37
\ms $pc^{-2}$ \\
\noindent Reference: (1) van den Bergh 1999; (2) Drimmel \& Spergel
2001; (3) Hammer et al. 2007; (4) van der Kruit 1986; (5) Tamm et
al. 2007; (6) Sackett 1997; (7) Kent et al. 1991; (8) Barmby et al.
2006; (9) Walterbos \& Kennicutt 1988; (10) Xue et al. 2008; (11)
Sakamoto et al. 2003; (12) Tempel et al. 2007; (13) Carignan et al.
2006; (14) Dehnen \& Binney 1998; (15) Geehan et al. 2006; (16)
Zheng et al. 2001; (17) Worthey et al. 2005; (18) Freudenreich 1998;
(19) Hiromoto et al. 1983; (20) Boissier \& Prantzos 1999; (21)
Fraternali 2009; (22) Williams 2003a; (23) Blitz et al. 1999; (24)
Holmberg \& Flynn 2004; (25) Klypin et al. 2002; (26) Widrow et al.
2003; (27) Dame 1993; (28) Kulkarni \& Heiles 1987; (29) Nieten et
al. 2006; (30) Dame et al. 1993; (31) Koper et al. 1991; (32)
Deharveng et al. 2000; (33) Daflon \& Cunha 2004; (34) Rudolph et
al. 2006; (35) Smartt et al. 2001; (36) Trundle et al. 2002.  \\

\end{table*}

\subsection{Gas and SFR Profiles}

The present-day profiles of gas and star formation  provide strong
constraints on models of the chemical evolution of a galactic disk.
In the case of the Milky Way, relevant observational data have been
collected in Boissier \& Prantzos (1999) and we adopt those data in
this work (Fig. 1, left panels). The gaseous profile is
characterized by a broad peak at a galactocentric distance $\sim4-5$
kpc (due to the ``molecular ring'' present at this distance) and the
SFR profile is also concentrated towards the inner disk. The total
gas mass is estimated to $\sim$7$\times$10$^9$ \ms and the total
star formation rate is SFR$\sim1-3$\msyr (e.g. Boissier \& Prantzos
1999, and references therein).

\begin{figure}[!t]
  \centering
  \includegraphics[width=0.48\textwidth]{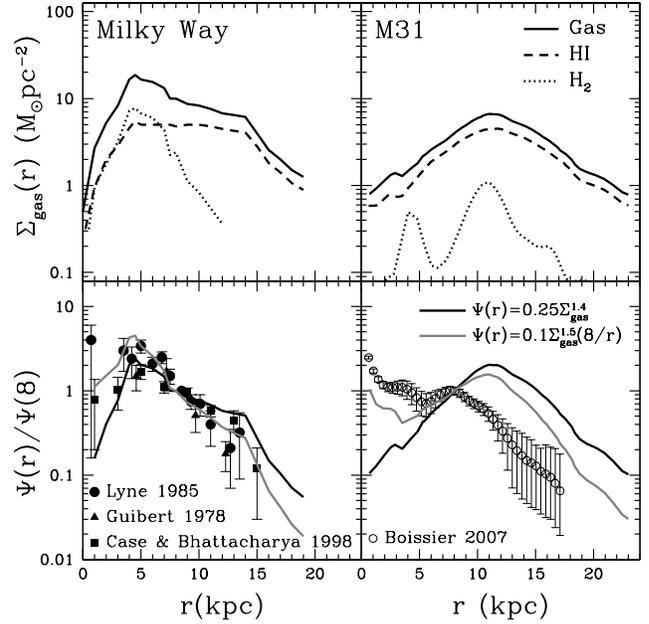}\\
  \caption{Observed profiles of the gas and star formation rate for
  the Milky Way and M31 disks. Upper panel: current surface density
  profiles of HI, molecular gas and total gas in the Milky Way
  and M31. Lower panel: observed estimations of SFR of the Milky Way
  (filled symbols, from Boissier \& Prantzos 1999) and M31 disks
  (open symbols from Boissier et al. 2007). The lines in the lower
  panels are the results calculated according to the star formation
  law given in the right panel.}
  \label{Fig:GasSFRobserved}
\end{figure}

For M31, the observed radial profiles for HI and H$_2$ gas surface
densities (Berkhuijsen 1977; Walterbos 1986; Koper et al. 1991;
Loinard et al. 1999) allow us to establish the radial gas profile
displayed in Fig.~\ref{Fig:GasSFRobserved} (right top). It is also
characterized by a broad peak, located at a galactocentric distance
twice as large as in the case of the MW. The HI profile from these
studies is also consistent with the one recently measured by Chemin
et al. (2009).

The star formation rate in several regions of M31 disk has been
carefully measured by both ground base photometry and Hubble Space
Telescope (Bellazzini et al. 2003; Williams 2002, 2003a,b; Brown et
al. 2006; Olsen et al. 2006). The current total SFR for M31 disk is
estimated to be 0.4 $\sim$1 \msyr (Williams 2003a,b; Barmby et al.
2006), i.e. less than half of the value in the Milky Way disk;
this shows that M31 is currently a rather quiescent galaxy.

Thanks to the GALEX UV satellite, it is now possible to obtain SFR
radial profiles for a number of local galaxies derived not from
H$\alpha$ data, but from the UV continuum (Boissier et al. 2007). In
Fig.~\ref{Fig:GasSFRobserved}, we show  the adopted gas (upper
panels) and SFR (lower panels) profiles of the Milky Way and M31
disks. From this figure, we see that the two spirals show quite
different properties in their gas and SFR profiles in the inner part
of the disk (between 3 and 7 kpc). The Milky Way has more gas in
this region, while M31 has most of its gas outside that region.

The total gas mass of the disks is obtained by integrating the gas
profiles from Fig.~\ref{Fig:GasSFRobserved} for both galaxies,
starting from the inner disk boundary (assumed to be at the bulge
radius $r_b\sim2.5$ kpc for the MW and $r_b\sim5$ kpc for M31)
outwards. We find M$_{gas,MW} \sim$ 7$\times$10$^9$ \ms and
M$_{gas,M31} \sim$ 6$\times$10$^9$ \ms (average value, taking
uncertainties into account). Since the M31 disk is twice as massive
as the Milky Way disk, its global gas fraction is about 1/2 of that
of the Milky Way disk (0.09 vs 0.19, respectively). This implies
that  M31 disk had an overall higher star formation efficiency than
the Milky Way (assuming that they have similar ages).

UV studies of star formation with GALEX showed that, in general, the
correlation between SFR and gas surface density is compatible with
empirical Kennicutt (1998a,b) SFR laws, with some scatter in the low
surface density side. But as Boissier et al. (2007) show, this
correlation fails for some individual galaxies, and especially for
M31 (see their Fig.6). In the lower two panels of
Fig.~\ref{Fig:GasSFRobserved}, we also show the expected behaviour
of two different SFR laws. The first depends only on gas surface
density, according to:
\begin{equation}\label{eq:sfrKennLaw}
    \Psi(r)=0.25~\Sigma_{gas}^{1.4}(r)
\end{equation}
from obsrvational data of Kennicutt (1998b, hereafter KS Law). The
second depends on both gas surface density and radius and is
motivated by the idea that star formation is induced by spiral waves
moving around a rotating disk (e.g. Wyse \& Silk 1989; see also Sec.
3.2):
\begin{eqnarray} \label{eq:sfrBP00}
\Psi(r)\propto\frac{\Sigma_{gas}^{n}(r)}{r}
\end{eqnarray}

It turns out that, in the case of the MW disk, observed gas and SFR
profiles fit well both SFR laws; as a result, the total star SFR is
also readily reproduced. But for M31, none of the SFR laws brings
agreement between observed gas and SFR profiles: there is a great
difference between theoretical expectations and observations in the
inner part of the disk (with little gas but, curiously,  high
observed SFR). As a result, any attempt to fit the SFR of the inner
disk with one of the aforementioned SFR laws will lead to an
overestimate of the SFR in the outer disk.

\begin{figure}[!t]
  \centering
  \includegraphics[width=0.48\textwidth]{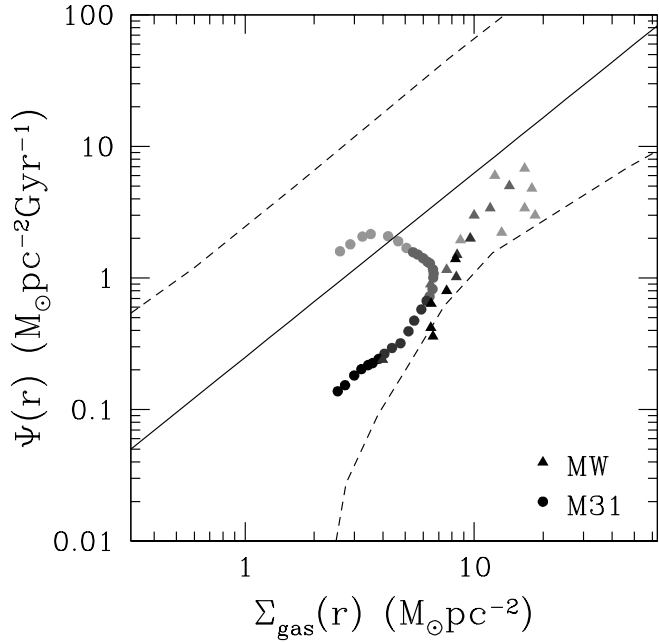}\\
    \caption{Relationship between observed gas surface density and
    star formation rate surface density for M31 (filled circles)
    and Milky Way disks (filled triangles).
    Full line is the classical Kennicutt law (equation(\ref{eq:sfrKennLaw})).
    Regions between two dashed lines indicate the results for a number
    of local galaxies observed by GALEX (from Boissier et al. 2007).
    The M31 shows an untypical path in its inner region, while the Milky
    Way shows a rather normal behaviour.}
  \label{Fig:SFRbyGas}
\end{figure}

In order to further demonstrate the different behavior of local star
formation rate in M31 and Milky Way disks, we plot the relationship
between the observed gas surface density and star formation rate
surface density in Figure~\ref{Fig:SFRbyGas}. Also we show the
general trend of the SFR with gas surface density (within  dashed
curves) for a number of nearby galaxies (from Boissier et al. 2007).
It is clear that M31 has a peculiar behaviour compared to both the
Milky Way and other local galaxies, especially in its inner region,
where high star formation rate corresponds to low gas amounts. In
the range 7-11 kpc the  SFR decreases when the gas surface density
increases, contrary to the classical SFR law. Then, beyond 11 kpc,
both SFR and gas amount decrease with radius, roughly following the
Kennicutt law.

If observations of SFR in M31 are not heavily distorted by incorrect
extinction corrections, then one concludes that the current SFR in
that galaxy does not obey one of the classical star formation laws
(Schmidt, Kennicutt, or some modified form of them). Perhaps, star
formation in M31 is (or has been) perturbed by some external event,
e.g. a major recent encounter with a galaxy of the Local group.
Indeed, observations of a two-ring-like structures by Block et al.
(2006) are interpreted as due to a  nearly central head-on encounter
with a companion galaxy (probably M32) about 200 Myr ago. If this is
indeed the case, then the present day SFR profile of M31 cannot be
used as a constraint on the chemical evolution model, since the
perturbation induced in the gaseous disk by the collision most
probably affected for (at least) one orbital time the SFR in M31.
Time-intergrated observables, like e.g. the total stellar profile or
the abundance profile (and, to a lesser extent, the gaseous profile)
certainly remain valid constraints.

\subsection{Disk Abundance Gradients}

Abundance gradients are an essential ingredient in an accurate
picture of galaxy formation and evolution (Boissier \& Prantzos
1999; Hou et al. 2000; Chiappini et al. 2001; Hou et al. 2002;
Cescutti et al. 2007; Magrini et al. 2009; Fu et al. 2009). The
existence of abundance gradients along the MW disk has been
established in the past twenty years using different tracers (Hou \&
Chang 2001). However, the magnitude of that gradient is still
subject to debate. Thus, oxygen or/and iron abundance gradients of
about $-0.06 \sim -0.07$ \dexkpc are obtained by using tracers as
HII regions and B stars (Rudolph et al. 2006 and references
therein), planetary nebulae (Maciel et al. 2006; Maciel \& Costa
2008) and open clusters (Chen et al. 2003,2008). However, values
about 40\% smaller are obtained by using those same tracers, e.g.
Deharveng et al. (2000, with HII regions), Daflon \& Cunha (2004,
using several tracers) and Andrievsky et al. (2004, with Cepheids).

The situation of the abundance gradient in the disk of M31 is also
far from clear. Early observations (Dennefeld \& Kunth 1981; Blair
et al. 1982) used HII regions and supernova remnants. A value of
dlog(O/H)/dr= $-0.06\pm 0.034$ \dexkpc  was derived using nebular
emission line ratios by Galarza et al. (1999).  The main uncertainty
comes from the empirical calibrations which are used to derive the
electronic temperatures in the nebular phase. By re-analyzing
earlier data from various authors, Trundle et al. (2002) find
smaller values for the oxygen abundance gradient, ranging from
$-$0.027 \dexkpc down to $-$0.013 \dexkpc. Furthermore, their
analysis of  five B-type supergiants covering the galactocentric
distance of 5-12 kpc leads to a negligible oxygen abundance gradient
of $-0.006 \pm 0.02$ \dexkpc; In contrast, they find a slightly more
significant gradient for Mg, of $-0.023 \pm 0.02$ \dexkpc.

In summary, the abundance gradient in M31 is very poorly known at
present. We adopt here a value of $-$0.017 \dexkpc, i.e. the mean
between the most extreme values of $-$0.027 \dexkpc and $-$0.006
\dexkpc found in the various analysis of Trundle et al. (2002). We
note that this value is substantially smaller (factor 3-4) than that
of the Milky Way disk ($-0.07$ \dexkpc), but only by a factor of 2
if the value of $-$0.04 \dexkpc \ is adopted for our Galaxy.
Finally, if we express the abundance gradients in terms of
corresponding scale lengths (dex/$r_d$), then the scaled gradient of
M31 disk is found to be two times smaller (equal) to the one of the
MW disk for the cases of $-$0.07 \dexkpc \  ($-$0.04 \dexkpc).

\begin{figure}[!t]
  \centering
  \includegraphics[height=9cm,width=9cm]{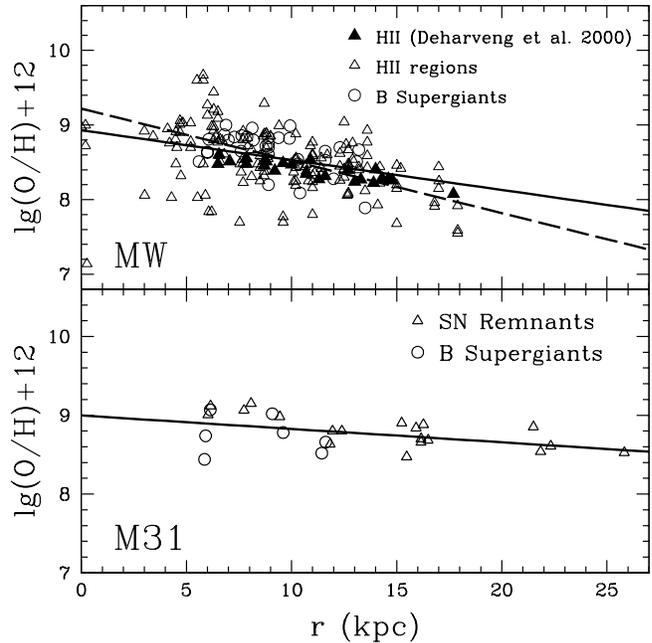}\\
    \caption{Observed oxygen abundance gradient in the Milky Way (top,
    data from Rudolph et al. 2006; Deharveng et al. 2000) and M31 (bottom,
    data from Dennefeld \& Kunth 1981; Blair 1982; Trundle et al. 2002).
    In the top panel, the two commonly referred values of $-$0.07 \dexkpc
    and $-$0.04 \dexkpc appear as {\it dashed} and {\it solid} line respectively.
    }
  \label{Fig:OxygenGradient}
\end{figure}

Figure~\ref{Fig:OxygenGradient} displays  the observed oxygen
abundance profiles in the Milky Way and M31 disks. In the case of
MW, two values for the gradient are shown, corresponding to $-$0.07
\dexkpc \ (circles) and $-$0.04 \dexkpc \ (solid line),
respectively.

\subsection{A Unified Description of the Milky Way and M31}

Table 1 summarizes the main observational features of the MW and M31
disks. The different sizes of the two major galaxies of the Local
group make difficult a direct comparison between their radial
profiles, thus giving no hints as to the physical ingredients
required for a successful simultaneous description of both disks. In
order to have a coherent picture, we attempt in this work a more
physical description, expressing all distances in terms of the
corresponding disk scale lengths.

\begin{figure*}[!t]
  \centering
  \includegraphics[height=14cm,width=18cm]{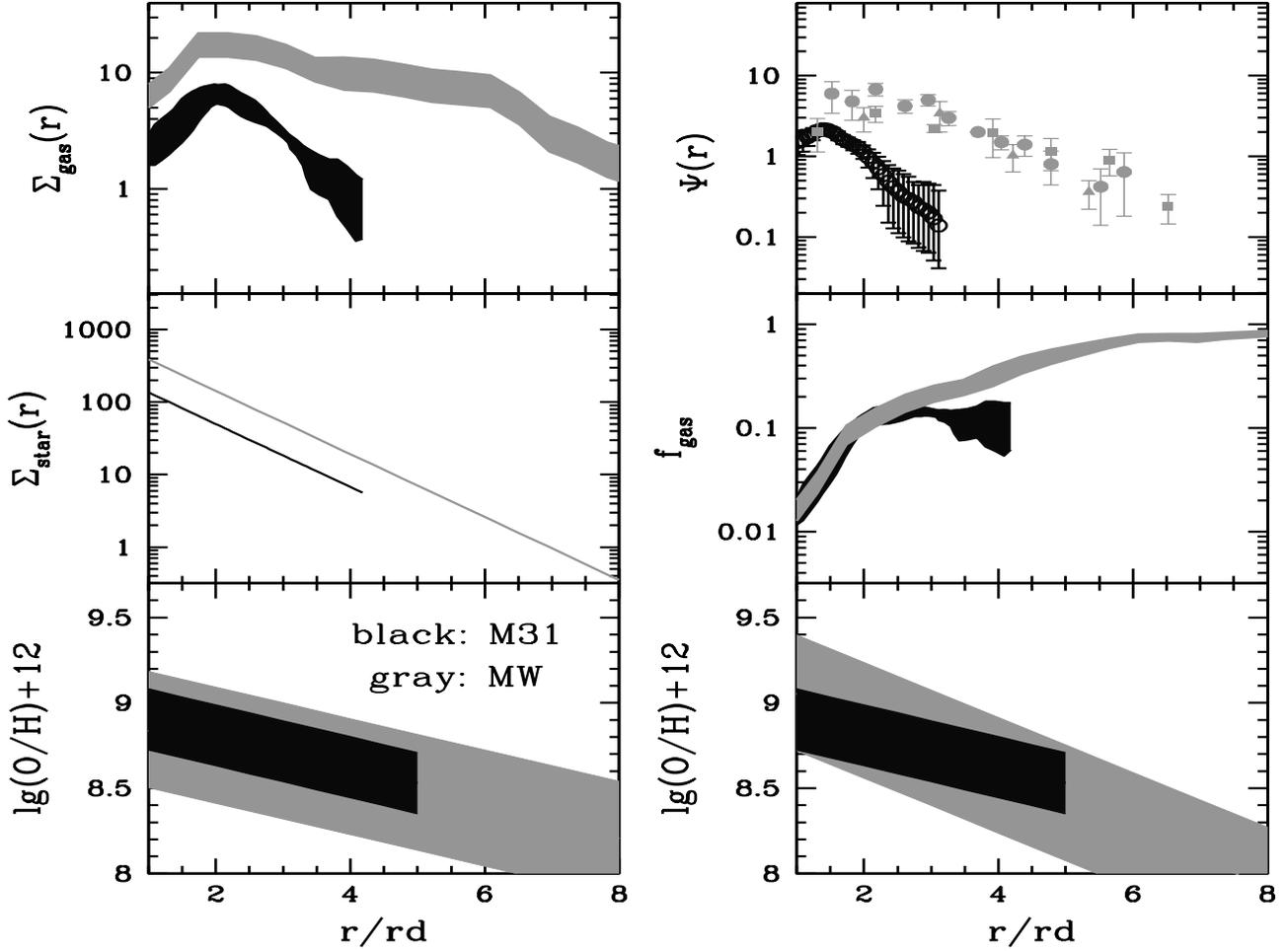}\\
    \caption{Observed profiles for Milky Way and M31 disks, with radius
    expressed in units of the corresponding scalelengths $r_d$. For the
    abundance gradient of the Milky Way disk, two sets of values are
    plotted, one is $-$0.07 \dexkpc (bottom right panel), the other is
    $-$0.04 \dexkpc (bottom left panel). In the latter case, MW and M31
    have similar abundance gradients, when expressed in dex/$r_d$. Shaded
    areas are the typical observed scatter. }
  \label{Fig:scaleobs}
\end{figure*}

In Fig.~\ref{Fig:scaleobs}, we plot the radial profiles of gas,
star, SFR, gas fraction, and the oxygen abundance gradients for the
two disks, using their scale lengths as distance units. It can be
seen that:

(1) The gaseous profiles (top left) are rather similar, in the sense
that they both display a broad peak at $\sim$2 scale lengths from
their centers. The MW has a more extended gaseous profile (in terms
of scale length).

(2) The Milky Way disk is more compact than M31, since it has a
higher stellar surface density at a given $r_d$ value (middle left).

(3) The profiles of scaled gas fractions (middle right) of the two
galaxies are quite similar in the inner disks. However, the overall
gas fractions of the two disks are quite different, with the MW
having a gas fraction twice higher than  that of M31.

(4) The scaled abundance gradients between two disks are similar if
we adopt the smaller reported value  for the MW disk (bottom left).

Thus, when the observed profiles are expressed in terms of scale
lengths, the two disks show some similarities in their properties.
One may hope then to describe both disks with a single chemical
evolution model, by varying as few as possible of the relevant
parameters. We describe such a model in the next section.

\section{The model}

In the case of the Milky Way, models with radially dependent infall
and star formation laws, forming the disk inside-out, are
generically used (Prantzos \& Aubert 1995; Boissier \& Prantzos
1999; Hou et al. 2000; Chiappini et al. 2001; Magrini et al. 2009).
Such models  reproduce several of the salient observational features
of the MW disk (including the abundance gradients), albeit with
different levels of success.

In this section, we present briefly our chemical evolution model
which is similar to the one adopted successfully in the past for the
Milky Way disk (see details in Boissier \& Prantzos 1999, 2000; Hou
et al. 2000).

\subsection{IMF and Stellar Yields}

The initial mass function (IMF) $\Phi(m)$ describes the mass
distribution of newborn stars and can be inferred from the observed
luminosity function on the basis of the mass-to-light ratio for
stars. Similar to our previous works, we adopt the IMF from the work
of Kroupa, Tout \& Gilmore (1993, KTG93), where some complex factors
(like stellar binarity, ages and metallicities, as well as
mass-luminosity and color-magnitude relationships) are explicitly
taken into account (Boissier \& Prantzos 1999; 2000; Hou et al.
2000).


Stellar yields are taken from Woosley \& Weaver (1995, WW95) for
massive stars, and from van den Hoek \& Groenewegen (1997, vdHG97)
for low and intermediate mass stars (mass from 0.8 to 8 \ms). They
are all metallicity dependent.


In order to account for the additional source of Fe-peak elements,
required to explain the observed decline of O/Fe abundance ratio in
the Milky Way disk (Goswami \& Prantzos 2000), we utilize the yields
of SNIa from the exploding Chandrashekhar-mass CO white dwarf models
W7 and W70 of Iwamoto et al. (1999). These are updated versions of
the original W7 model of Thielemann et al. (1986), calculated for
metallicities Z = \zs (W7) and Z = 0 (W70), respectively.

\subsection{Infall Rate and Timescale}

We assume that the MW and M31 disks are progressively built up by
infall of primordial gas cooling down from their dark haloes. The
form of the time dependence of the infall rate is unknown at
present. In Prantzos \& Silk (1998), an asymmetric Gaussian infall
rate was assumed, on the basis of dynamical arguments. However,
usually simpler parametrizations are adopted, i.e. infall rate is
exponentially decreasing in time:


\begin{equation} \label{eq:infall}
f(t,r) \  = \ A(r) \ e^{-t/\tau(r)}
\end{equation}
where $A(r)$ is a normalizing function and can be obtained by:
\begin{equation}\label{eq:Ar}
    \int_0^{t_g}A(r)\cdot e^{-t/\tau(r)}dt=\Sigma_{tot}(r,t_g)
\end{equation}
where $\Sigma_{tot}(r,tg)$ is the current total mass profile and
$\tau(r)$ is the infall time scale which is radially dependent. In
the Milky Way disk, the characteristic infall time scale in the
solar neighborhood ($R_{\odot MW}$ = 8~kpc)is $\sim$7 Gyr (Chiappini
et al. 1997; Boissier \& Prantzos 1999; Chang et al. 1999, 2002), in
order to reproduce the local G-dwarf metallicity distribution.

The radial dependence of the infall time scale for the MW disk is
given by $\tau_{MW}(r)=b~r/r_d$, where $r_d$ is the scale length and
$b$ is a free parameter. Positive values of $b$ imply an inside-out
formation of the disk and we adopt here $b$=2.5, which leads to
formation time scales of $\sim$2 Gyr for the  inner disk and
$\sim$10 Gyr for the outer disk.

In the case of M31, we adopt the prescription used in Boissier \&
Prantzos (2000), according to which the infall time scale is assumed
to be correlated with both surface density and galaxy mass:
\begin{equation}\label{eq:tau}
    \tau^{-1}(r)=\tau_{MW}^{-1}(r)+0.4(1.0-\frac{V_{C}}{220})
\end{equation}
where $V_{C}$ (in km~s$^{-1}$) is the flat rotational velocity for
the galaxy disk and $\tau_{MW}(r)$ (in Gyr) is the infall time scale
for the Milky Way disk.  According to Hammer et al. (2007), $V_{C}$
for M31 is about 226 km~s$^{-1}$, i.e. the same as that of Milky Way
disk. Therefore, our adopted prescription leads to similar infall
time scale laws for both disks.

\subsection{Star Formation}

The star formation rate  remains the major unknown in chemical
evolution studies. Kennicutt (1998a,b) found that the global SFR of
disks and circumnuclear starburst galaxies is correlated with the
local gas density  over a large range in surface density and SFR per
unit area, spanning 5 orders in magnitude. Over that range, the
empirical SFR vs gas surface density relation can be fitted by a
simple power law with index $n \sim 1.4$. Kennicutt (1998a,b) also
found that the data can be fitted equally well as a function of the
local dynamical time scale, $\tau_{dyn}$: $\Psi
\propto\frac{\Sigma_{gas}}{\tau_{dyn}} \propto\Sigma_{gas}
 \Omega$, where $\Omega$ is the rotation speed of the gas. Since
$\Omega \sim V(r)/r$, the SFR could be expressed as:
\begin{equation}
    \Psi(r) \ \propto \  \Sigma_{gas}\ {{V(r)}\over{r}}
\end{equation}
where $V(r)$ is the circular velocity at radius $r$. Since spiral
galaxies display $V(r)\sim$ constant, one gets a modified
Kennicutt-Schmidt law (hereafter M-KS law), as suggested on
theoretical grounds in Wyse \& Silk (1989, see also Prantzos \&
Aubert 1995).

Boissier \& Prantzos (1999) adopted the index $n$ of the M-KS SFR
law to be $n=1.5$ on an empirical basis, in order to fit the present
day profiles of the MW SFR (Fig. 1, bottom left). They also adopted
this M-KS law in subsequent models for external spirals, which can
successfully reproduce most of the chemical and photometric
properties of disk galaxies (Boissier \& Prantzos 2000; Boissier et
al. 2001) and in particular the observed abundance gradients
(Prantzos \& Boissier 2000). In a recent study, Fu et al. (2009),
 have used both KS law and M-KS law to predict the time
evolution of Galactic disk abundance gradient. By comparing the
model predictions with the observed results from open clusters and
planetary nebulae with different ages, it is concluded that by
adopting the M-KS law, model results are more consistent with the
observed evolution of abundance gradient. Therefore, we will adopt
this M-KS law for Milky Way and M31 disks:
\begin{equation} \label{eq:sfrB}
    \Psi(r) = \alpha~\Sigma_{gas}^{1.5}(\frac{r_{eq\odot}}{r})
\end{equation}
The coefficient $\alpha$ is related to the star formation
efficiency. All other things being equal, it appears that the star
formation efficiency in M31 has to be at least twice as high as in
the MW, since its observed gas fraction is twice as small (Table 1
and discussion in Sec. 2.2). We shall see indeed in the next section
that such a larger $\alpha$ is required in order to fit the M31
data.

\begin{table}[!t]
\noindent Table 2. Model parameters \\
\begin{tabular}{lll}
\hline \hline
 General                       &   Prescription               &  Parameter \\
\hline
 IMF                           &   KTG1993                    &            \\
 Mass limits                   &   (0.1-100) \ms              &            \\
 SFR                           &   $\alpha$$\Sigma_{gas}^{1.5}(r_{eq\odot}/r)$ &  $\alpha$  \\
 Stellar yields                &   vdHG97,WW95                &            \\
 Metallicity of infall gas     &   Z$_{f}$ = 0                &            \\
 Infall time scale             &   $\tau(r) = b \ (r/r_d)$    &   $b$        \\
 Age of disk (Gyr)             &   13.5                       &            \\
\hline
Individual                     &   Milky Way                  &   M31         \\
\hline
 Scale length $r_{d}$ (kpc)     &   2.3                       &   5.5           \\
 Equivalent $r_{eq\odot}$(kpc)  &   8.0                       &   19.0    \\
 Total disk mass ($10^{10}$\ms) &  5.0                        &   7.0           \\
 V$_{rot}$(\kms)                &  220                        &   226           \\
\hline \hline
\end{tabular} \\
\end{table}

\begin{figure*}[!t]
  \centering
  \includegraphics[angle=-90,width=18cm]{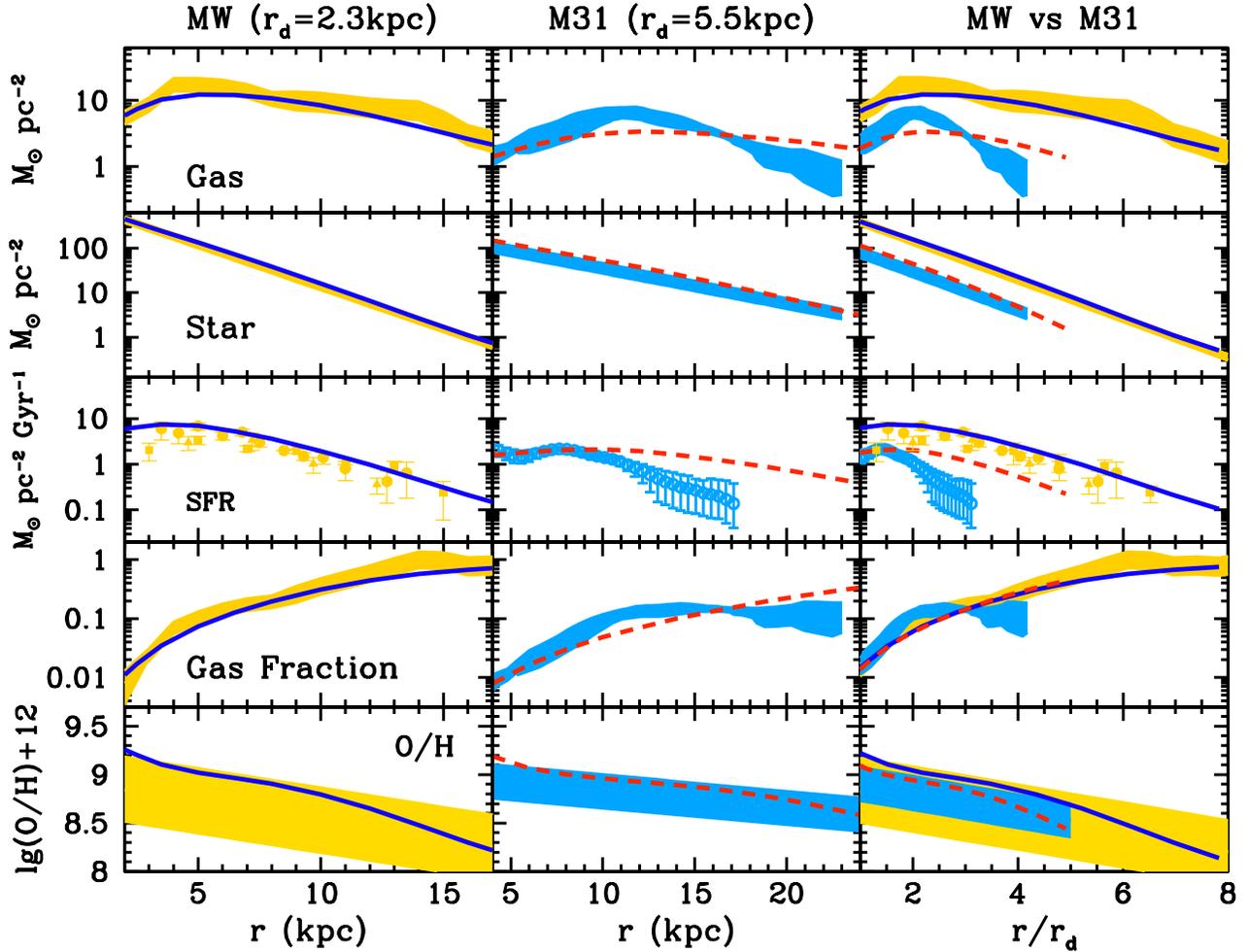}\\
  \caption{Current profiles of gas, stars, SFR, gas fraction and oxygen abundance
  (from top to bottom) for MW and M31. In the {\it left} and {\it middle} panels,
  profiles for MW and M31, respectively, are expressed in terms of physical radius
  $r$ (in kpc); in the {\it right} panels, profiles for both disks are expressed
   in terms of normalised radius $r/r_d$. Observations are presented
   as yellow shaded ( for MW with $-0.04 dex/kpc$ ) or blue shaded ( for M31 )
   areas and model results by solid (for the MW) and dashed (for M31) curves,
   respectively.
  }
  \label{Fig:Profiles}
\end{figure*}

\section{Model Results and Comparison with the Milky Way and M31 Disks}

We run our simulations with the parameters of Table 2 for MW and M31
disks. Notice that we adopt  the same model parameters for both
galaxies (hence, attempting to describe them in a unified
framework), except for: (i) the small difference in the infall rate
from Eq. (\ref{eq:tau}), which makes M31 slightly older than the MW
and (ii) the star formation efficiency parameter $\alpha$, assumed
here to be twice as large for M31 than for the MW, i.e. $\alpha_{MW}
= 0.1$ and $\alpha_{M31} = 0.2$.

\subsection{Radial Profiles}

In Fig.~\ref{Fig:Profiles}, we show the model predictions of the
radial profiles for gas, stars, SFR, gas fraction and oxygen at time
$t$=13.5 Gyr and we compare them with observational data. The first
two columns display results for MW and M31, respectively, as a
function of radius $r$ expressed in kpc. The third column presents
the same results in a common scale of normalized radius $r/r_d$ for
both galaxies; this allows to better visualize the similarities and
differences between the two disks.

The main results of the comparison with observations can be
summarized as follows:

1) In both cases, exponential disk profiles are obtained by
construction, since most of the infalling gas (the radial profile of
which is normalised through Eq.(~\ref{eq:Ar})) is  turned into
stars.

2) The model gaseous profiles go through a broad maximum, obtained
at the observed position, approximately at two scale lengths from
the galactic centers. This maximum is obtained in the models through
the radial dependence of the SF efficiency (being greater in the
inner disk, it produces a gas fraction profile $f_{gas}(r)$
increasing with radius, see next paragraph) and the total surface
density profile $\Sigma_{tot}(r)$ which decreases with radius (by
construction). The gaseous profile being the product of the two
($\Sigma_{gas}(r)=f_{gas}(r) \Sigma_{tot}(r)$), the resulting curve
goes through a maximum, and within our unified scheme this happens
at $\sim$2 $r_d$.

3) The gas fraction profile decreases monotonically inwards, in
perfect agreement with observations for the MW  and in fair
agreement for the inner disk of M31. Only for the outer disk of M31
the model predicts slightly higher than observed gas fractions. We
notice that, in terms of normalized radius $r/r_d$, the gas fraction
profiles of the two disks are very similar, which explains their
successful description by our unified model. Notice that, in terms
of {\it physical radius}, M31 has a smaller gas fraction than the MW
at a given $r$, which explains the need for a higher SF efficiency
in that case. The situation is less satisfactory in the outer disk
of M31, where the gas fraction is overestimated by our model.

4) Our model predicts correctly the present day SFR profile of MW,
but fails completely in the case of M31, and in particular in the
outer disk of M31. As already noticed (Sec. 2.2) the observed SFR vs
gas relationship in M31 cannot be fit by any form of the KS laws.
Our result reflects just this impossibility. As already argued (last
paragraph of Sec. 2.2), we believe that the observed SFR is affected
by recent perturbations of the gaseous disk of M31, e.g. the
collision with a nearby galaxy suggested by Block et al. (2006).

5) The resulting abundance gradients are compatible with
observations for both MW and M31. In fact, the predicted abundance
profile of MW is somewhat steeper than in the case of M31. This is
not a surprise since the two disks do not have the same scale. At
the same distance from the galactic center, they have different gas
  amounts and SFR. As a result of its larger scalelength, the current
  gas is more widely spread in M31 than in the Milky Way, with a
  resulting gas fraction rising less steeply in Andromeda, and
  correspondingly a flatter abundance gradient.  When we express the
  model abundance gradient in terms of their scale length, we obtain
  similar value for the two disks. In any case, taking into account
all the uncertainties mentioned in Sec. 2.3, we consider the overall
agreement as satisfactory. Further observations will hopefully
establish the true abundance profiles of MW and M31 with greater
accuracy, perhaps pointing to some different prescriptions for our
unified model.

\subsection{Infall and Star Formation History }

\begin{figure}[!t]
  \centering
  \includegraphics[height=10cm,width=9cm]{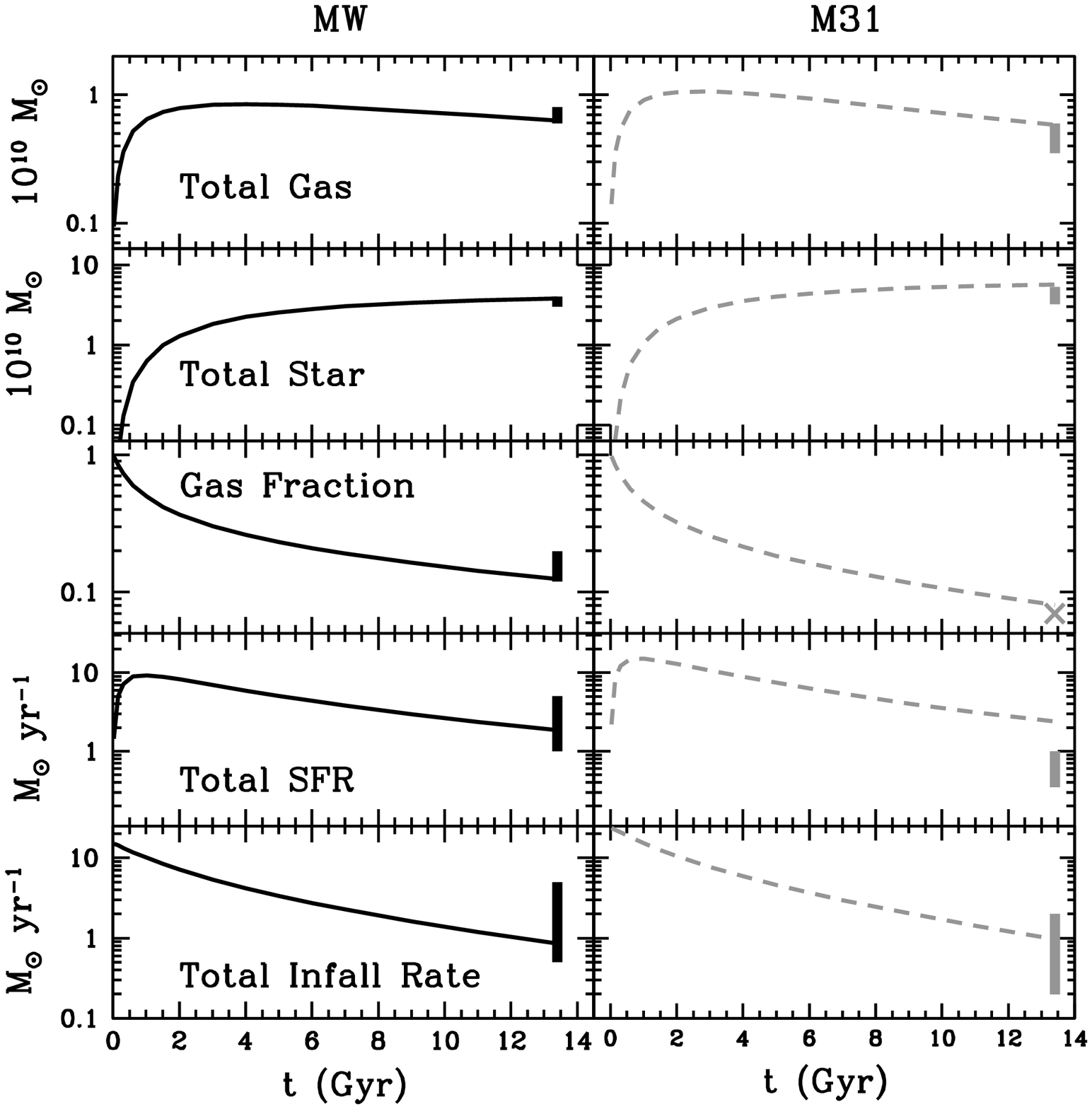}\\
  \caption{Time evolution of global gas, star, gas fraction, infall
  rate and SFR in M31 and Milky Way disks. The disk parameters are
  given in Table 2. And the coefficients $\alpha$ of SFR in Table 2
  for the Milky Way and M31 disks are $\alpha_{MW} = 0.1$, $\alpha_{M31} = 0.2$,
  respectively. Infall time scale is $\tau(r) = 2.5~r/r_d$.
  Bar in the right of each plot gives the observed estimations.
  It can be seen that the model predicts two much present SFR
  for M31 disk. }
  \label{Fig:InfallSFRtime}
\end{figure}

In Fig.~\ref{Fig:InfallSFRtime}  we show the evolution of the total
amount of gas, stars, SFR and infall rate, and of the gas fraction
for the disks of MW and M31. Reasonable agreement with observations
is obtained for all those quantities in the case of MW disk; this
agreement results from the adopted normalization of the total disk
mass and the adopted star formation efficiency. In the case of M31,
the model predicts current global SFR$\sim$2.0 \ms $yr^{-1}$, which
is substantially larger than  observational estimates (Williams
2003a,b; Barmby et al. 2006). In view of the discussion in Sec. 2.2,
we do not consider this discrepancy as significant: recent star
formation in M31 may have been considerably perturbed by external
effects (i.e. collision with another galaxy), unaccounted for in our
model. We also notice that the current infall rate is poorly
constrained in MW and virtually unconstrained in M31. In the case of
M31, Thilker et al. (2004) found that there exists an extensive
population of HI clouds in the outskirts of the galaxy. The values
displayed in Fig. 6 are in the range of 0.2-2 \ms $yr^{-1}$, the
former being the typical value inferred from observations of
accreting cold gas in disks (Sancisi et al. 2008) and the latter
from a simple theoretical argument, namely that such values are
required to maintain a quasi-constant SFR over disk history for
MW-size disks.

\subsection{Metallicity Distributions}

\begin{figure}[!t]
  \centering
  \includegraphics[height=10cm,width=9cm]{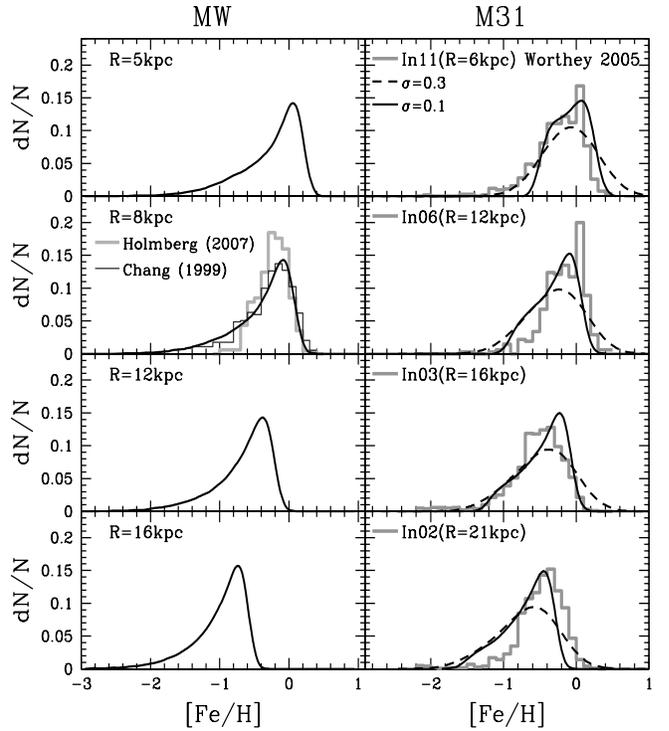}\\
  \caption{The metallicity distribution functions of various regions
  in the Milky Way and M31 disks. The observed data in the solar neighborhood
  of the Milky Way disk come from Holmberg et al. (2007) and Chang et al.
  (1999), and the data of M31 come from Worthey et al. (2005).
  The model predictions are plotted as smooth curves,
  after convolution with a Gaussian error function with $\sigma$ = 0.10 dex
  (full lines) and 0.3dex (dashed lines). For M31, we construct
  metallicity distributions for the age-range of RGB stars observed by
  Worthey et al.(2005) (stellar ages between 6 and 12 Gyr). }
  \label{Fig:MDFs}
\end{figure}

In this subsection, we compare the model metallicity distribution
functions (MDFs) with currently available observations in various
regions of the Milky Way and M31 disks. The model results correspond
to main sequence stars with lifetimes $\tau>$10 Gyr, and they have
been convolved with Gaussian error functions with $\sigma$ = 0.1 dex
(Fig.~\ref{Fig:MDFs}).

For the Milky Way disk, the observed data in the solar neighborhood
are from the GK survey (Holmberg et al. 2007, who revised estimates
in Nordstr\"{o}m et al. (2004)), which includes ages, metallicity
and kinematic properties for about 14000 F and G dwarfs. MDFs in
other regions are not available at present, but future surveys e.g.
SDSS/SEGUE (Ivezi\'{c} et al. 2008) and China's LAMOST project (Zhao
et al. 2006), are expected to provide information on those regions
as well. As expected (from the adopted infall rate)  our model fits
rather well, albeit not perfectly, the local MDF. Notice that we
compare to the data of Holmberg et al. (2007) corrected for the
scale height of sellar populations (dashed curve in their Fig. 22,
right panel): indeed, our results concern the full extent of the
so-called ``solar cylinder'' at 8 kpc from the Galactic center,
while local surveys are complete only within a limited volume
centered on the Sun. However, the corrections in Holmberg et al.
(2007) are made after some assumptions are made about the star
formation history of the local disk, which is not necessarily the
same as the SF history in our model. Thus, it should not be
surprising that the fit is not perfect.

In the case of M31, data are available for the MDF in various places
along its disk (Bellazzini et al. 2003; Worthey et al. 2005; Chapman
et al. 2006). Right panel histograms are results from Worthey et al.
(2005), who observed 11 regions from the inner regions to the outer
disk along the major axis of M31. The median abundances in each
observed field increase steadily from the inner to the outer disk.
The mean stellar metallicity is $\sim$0.2 dex lower than the
gas-phase abundance (Fig.~\ref{Fig:OxygenGradient}) in the same
location.

In the right panels of Fig.~\ref{Fig:MDFs}, we present also the
model predictions for the MDFs in the same  radial positions as the
data available for M31. Our results are in broad agreement with
observations and they reproduce the decrease of mean stellar
metallicity with radius, as a consequence of the star formation and
infall schemes adopted for the disk. We conclude then that, to a
first approximation, M31 evolved inside-out, as expected for a
normal spiral. Despite this, rather satisfactory agreement, between
the data and our model, we would like to emphasize the need for more
data on the MDFs of M31 as a function of radial position in order to
further constrain the evolution of its disk.

It should be noted that when we plot the model predicted MDFs for
M31 disk, we have assumed an error about 0.1dex based on the Worthey
et al. (2005). This assumption is self-consistent with observations
since we have used the data from Worthey et al. (2005). But this
adopted error on the photometric metallicity may be too small as we
know that even for the situation of the halo, where the age spread
should be smaller than in the disk (and associated uncertainties
lower) a comparison of spectroscopic and photometric metallicities
for RGB stars in M31 should scatter about $\pm$0.3 dex (see Kalirai
et al. 2008). Therefore, we also plot the model MDFs for M31 disk
with photometric error of 0.3dex by dashed lines in Fig. 7. As
expected, the model predicted MDFs are wider than the observed
distributions in this case. But the peak position is roughly the
same.

We emphasis that while calculating the models, the results are for
disk evolution with full star formation history, that is, includes
all stars in the disk with all ages. Worthey et al.\,(2005) did not
discuss the age spread in details, however they claimed that the age
spread for their RGB stars is about 6-12Gyr (section 2, last
paragraph in Worthey et al.). Therefore, we construct the
metallicity distributions for M31 disk with the age-range of RGB
stars observed by Worthey et al. (stellar ages between 6 and 12 Gyr)
and compare them to their observations in Fig. 7.

On the other hand, Koch et al. (2005) found that in the Carina dSph,
there is a larger age spread ( from 2Gyr to more than 11Gyr ), while
its color-magnitude diagram shows a narrow RGB distribution. The
large age spread means that Carina dSph must have undergone various
episodes of star formation process. This is different from the
smaller age spread reported by Worthey et al. (2005). We think this
uncertainty calls for a more work on the observational side about
the metallicity distribution of M31 disk.

\section{Discussion}

Early works on simultaneous modeling of MW and M31 (Diaz \& Tosi
1984) found some similarities between the evolutionary properties of
the two disks, but they were performed at an epoch where scarce
observational data provided little constraint to the models (for
instance, Diaz \& Tosi 1984 compared their model to M31 data
available only in the 5-11 kpc region). A more detailed comparison
to observations is made in Molla et al. (1996), who use a
multi-parameter model and reproduce successfully several features of
the M31 disk.

The recent work of Renda et al. (2005) focuses on MW and M31,
benefits from a larger data set and presents some similarities to
our work. The disks are constructed inside-out by slow infall and
the adopted SFR is $\Psi = \alpha \Sigma_{Gas}^2/r$, i.e. with an
exponent $n$=2 instead of 1.5 in our case. By adopting exactly the
same SFR law for MW and M31, Renda et al. (2005) find that the
gaseous profile is over predicted in M31 (their model M31a), hence
the need to increase their SF efficiency $\alpha$ by a factor of 2
in order to improve the fit to the data (their model M31b). Had they
noticed the lower gas fraction in M31, they would have anticipated
the problem (see our discussion in Sec. 3.3).  Our  models agree
both in the conclusion for a higher SF efficiency in M31 compared to
the MW as well as on the resulting abundance gradients (smaller in
the case of M31), when radius in all radially dependent terms is
expressed in e.g. kpc. For some unclear reason, our model fits
better the gaseous profile of M31 (perhaps, because of our smaller
exponent $n$=1.5 in the adopted SFR). Finally, both our model and
theirs fail in the outer disk of M31. It is hard to push the
comparison further, since Renda et al. (2005) do not provide SFR and
stellar or  gas fraction profiles. The latter are in fact, mandatory
in any work on chemical evolution, since they constrain, more than
anything else the combined history of star formation and infall.

Despite their simplicity (independently evolving rings, no
cosmological framework), models such as the one presented here can
provide some interesting physical insights to the evolution of MW
and M31, based both on their successes and their failures. The
success in reproducing simultaneously the profiles of gas, SFR and
metallicity in the MW, as well as its global properties (gas
fraction, total SFR and colours, the latter being discussed in
Boissier \& Prantzos (1999)) implies that the overall history of the
Milky Way cannot have been very different from the one found here,
i.e. a slow, inside-out disk formation. However, similar solutions
may, perhaps, be obtained by some other combinations of SFR and
infall rate, i.e. the problem may well be degenerate, thus no firm
conclusions can be drawn on each one of those two key ingredients.

In the case of M31, it is clear that a higher star formation
efficiency is required, as deduced from its gas fraction, smaller by
a factor of $\sim$2 than in the MW. This was already found by Renda
et al. (2005), while Hammer et al. (2007) went one step further, to
suggest that the MW is a particularly ``quiescent'' disk galaxy (for
its mass) and M31 may be closer to an average large spiral. This
``quiescence'' of the MW may be due to its relative isolation, while
M31 may have undergone a larger number of (and/or more important)
interactions with neighboring  galaxies. Such a picture is in line
with the finding of Block et al. (2006), namely that M31 has
undergone a major interaction about 200 Myr ago; no such interaction
appears to have occurred in the case of the MW over the last
billions of years.

Our formalism allows us to describe in a unified framework the
properties of both the MW and M31, by using the same expression for
the radial dependence of the SFR in both cases. Such a description
is demanded by the similarity in the radial profiles of those two
disks, when they are expressed in terms of their respective scale
lengths (Sec. 2.4). However, it is not clear whether the higher SF
efficiency of M31 is due to an external factor (i.e. more
frequent/important interactions of that galaxy) or to an internal
one (e.g. its mass, as argued in Boissier \& Prantzos 2000).
Applying this formalism to other disk galaxies for which large data
basis are available (work in progress) will help to clarify the
situation.

On the other hand, the failure of both this work and Renda et al
(2005)  to reproduce satisfactorily the gaseous profile of M31, and
the fact that we over predict the global SFR of M31, as well as its
outer SFR profile, suggests that those properties are considerably
affected by recent interactions. Thus, they cannot be predicted by
such simple models (unless if more parameters are introduced). If
this is true, and if M31 is really closer to a typical disk (as
Hammer et al. 2007 suggest), then the cosmological framework will be
mandatory for the description of galactic disks; simple models, like
this one, will be able to describe successfully only the most
quiescent disks, such as the MW.

\section{Summary}

In this work, we study the chemical evolution of the disk of M31,
using a  model already applied to the study of the  Milky Way
(Boissier \& Prantzos 1999, Hou et al. 2000). We use an extensive
data set of M31 properties, including radial profiles of gas surface
density, gas fraction, star formation rate, oxygen abundances, as
well as metallicity distribution functions at different regions of
the disk.  In particular, the star formation profile of M31  is from
recent UV data of GALEX (Boissier et al. 2007). Our main purpose is
to see whether a simple chemical evolution model can successfully
describe the radial and global properties of both disks.

We first summarize and compare the  observational data (Sec. 2) for
the two galaxies. The disk of M31 is about 2.4 times larger  and 2
times more massive than the Milky Way disk, while its gas fraction
is approximately half of the one of the MW. All other things being
equal,this implies a higher average star formation efficiency for
M31. We find that the SF radial profile of MW is well described by
``standard'' SF laws, but not the one of M31 (Sec. 2.2). We
attribute the latter to a recent major perturbation of M31 by a
nearby galaxy, in line with the findings of Block et al. (2006). We
conclude that our model (which adopts such ``standard'' SF laws)
will fail to reproduce the observed SF profile of M31, and perhaps
also the gas profile.

We find that, when radii are expressed in terms of the corresponding
scale lengths, the two disks display very interesting similarities
in their radial profiles (Sec. 2.4). This concerns, in particular,
the gas fraction, the profile of which is quasi-identical inside the
innermost two scale lengths (Fig.~\ref{Fig:scaleobs}). Also, the
scaled abundance gradients of the two disks are quite similar {\it
if} we adopt for the MW the lower range of reported values (e.g.
Deharveng et al. 2000; Daflon \& Cunha 2004; Andrievsky et al. 2004;
Chen et al. 2008). Such a similarity was found in a sample of
external spirals and successfully described by the models of
Boissier \& Prantzos (2001), which cover a much larger range of
galaxian properties than the two disks studied here. We stress,
however, that the status of the MW abundance gradient, especially in
the outer part, is still very controversial: observations show that
it may not be described by a simple exponential (see e.g. Yong et
al. 2005 and Carraro et al. 2007 for open cluster abundances; and
Andrievsky et al. 2004 and Lemasle et al. 2008 for Cepheids). {\it
Assuming that} the scaled abundance gradients are similar in MW and
M31, we seek then a description of the radial properties of the two
disks within the framework of a single model, which we present in
Sec. 3.

Detailed calculations show that our unified model describes fairly
well all the main properties of the MW disk and most of those of
M31, provided its SF efficiency is adjusted to be twice as large in
the latter case (as anticipated from the lower gas fraction of M31).
The radial profiles of both MW and M31 are well described, albeit
less successfully in the case of M31. In particular, the model fails
to match the present SFR in M31, producing too large values in the
outer disk and globally. We attribute this failure to the fact that
M31 has been perturbed recently by a major encounter, as already
anticipated by the fact that the observed SFR profile of M31 does
not seem to follow any form of the Kennicutt-Schmidt star formation
law. On the other hand, the stellar metallicity distributions
measured along the disk of M31 reflect the integrated star formation
during the whole disk history and should not be affected by recent
events. Our model, where the bulk of Fe originates in SNIa,
reproduces rather well those distributions, from 6 to 21 kpc.

The unified description that we propose here for MW and M31, by
expressing their radial profiles in terms of the ``natural units''
(the corresponding disk scale lengths), offers valuable insights
into the evolution of those two disk galaxies and this may also be
the case for other spirals as well (work is in progress).

\acknowledgements This work is supported by the National Science
Foundation of China No.10573028, the Key Project No.10833005, the
Group Innovation Project No.10821302, and by 973 program No.
2007CB815402.

\def\cjaa{CJAA}

{}

\end{document}